\documentclass[11pt,fleqn]{article}

\usepackage{amsfonts,amssymb,latexsym,cite}
\usepackage{graphicx}
\usepackage{amsmath, amssymb}
\usepackage{epstopdf}
\usepackage{eurosym}

\usepackage{amsfonts,amssymb,cite}
\usepackage{graphicx}

\def\noi{\noindent}

\newcommand{\Title}[1]{\noi {{\Large\bf #1}}\\[1ex]}

\def\Aunames#1{\noi{\bf #1}}
\def\au#1{${}^{#1}$}
\def\Addresses#1{\medskip\noi \protect
	\begin{description}\itemsep -3pt {\it #1} \end{description}}
\def\adr#1#2{\item[${}^{#1}$]{\it #2}}

\newcommand{\Abstract}[1]{\vskip 2mm \begin{center}
        \parbox{16.4cm}{\small\noi #1} \end{center}\medskip}

\def\email#1#2{\footnotetext[#1]{e-mail: #2}\addtocounter{footnote}{1}}

\def\nqq{\hspace*{-2em}}
\def\nhq{\hspace*{-0.5em}}

\def\cm{\hspace*{1cm}}
\def\inch{\hspace*{1in}}

\def\Jl#1#2{#1 {\bf #2},\ }

\def\ApJ#1 {\Jl{Astroph. J.}{#1}}
\def\CQG#1 {\Jl{Class. Quantum Grav.}{#1}}
\def\DAN#1 {\Jl{Dokl. AN SSSR}{#1}}
\def\GC#1 {\Jl{Grav. Cosmol.}{#1}}
\def\GRG#1 {\Jl{Gen. Rel. Grav.}{#1}}
\def\JETF#1 {\Jl{Zh. Eksp. Teor. Fiz.}{#1}}
\def\JETP#1 {\Jl{Sov. Phys. JETP}{#1}}
\def\JHEP#1 {\Jl{JHEP}{#1}}
\def\JMP#1 {\Jl{J. Math. Phys.}{#1}}
\def\NPB#1 {\Jl{Nucl. Phys. B}{#1}}
\def\NP#1 {\Jl{Nucl. Phys.}{#1}}
\def\PLA#1 {\Jl{Phys. Lett. A}{#1}}
\def\PLB#1 {\Jl{Phys. Lett. B}{#1}}
\def\PRD#1 {\Jl{Phys. Rev. D}{#1}}
\def\PRL#1 {\Jl{Phys. Rev. Lett.}{#1}}

\def\al{&\nhq}
\def\lal{&&\nqq {}}
\def\eq{Eq.\,}

\def\beq{\begin{equation}}
\def\eeq{\end{equation}}
\def\bear{\begin{eqnarray}}
\def\bearr{\begin{eqnarray} \lal}
\def\ear{\end{eqnarray}}
\def\earn{\nonumber \end{eqnarray}}

\def\nnn{\nonumber\\ \lal }

\def\yy{\\[5pt] {}}
\def\yyy{\\[5pt] \lal }
\def\eql{\al =\al}

\def\dst{\displaystyle}
\def\tst{\textstyle}
\def\fracd#1#2{{\dst\frac{#1}{#2}}}
\def\fract#1#2{{\tst\frac{#1}{#2}}}
\def\Half{{\fracd{1}{2}}}
\def\half{{\fract{1}{2}}}

\def\e{{\,\rm e}}
\def\d{\partial}

\def\const{{\rm const}}

\def\mN{_\mu^\nu}

\def\sph{spherically symmetric}
\def\ssph{static, spherically symmetric}

\def\Schr{Schr\"odinger}

\def\rf{\eqref}
\def\eqn#1{\eq \rf{#1}}

\def\R{{\mathbb R}}

\def\da{\delta\alpha}
\def\db{\delta\beta}
\def\dg{\delta\gamma}
\def\df{\delta\phi}

\usepackage{color}

\begin{document}
\twocolumn[
\vspace{2cm}

\Title{On the instability of some k-essence space-times}

\Aunames{K. A. Bronnikov,\au{a,b,c,1}
	 J. C. Fabris,\au{c,d,2}
	 and Denis C. Rodrigues\au{d}
	 }

\Addresses{
\adr a {VNIIMS, Ozyornaya ul. 46, Moscow 119361, Russia}
\adr b {Institute of Gravitation and Cosmology, PFUR,
         ul. Miklukho-Maklaya 6, Moscow 117198, Russia}
\adr c {National Research Nuclear University ``MEPhI'',
	 Kashirskoe sh. 31, Moscow 115409, Russia}
\adr d {N\'ucleo Cosmo-ufes \& Departamentno de F\'{\i}sica, Universidade Federal do Esp\'{\i}rito Santo,
 	 Vit\'oria, ES, CEP 29075-910, Brazil}
	}

\Abstract
   {We study the stability properties of static, spherically symmetric
   configurations in k-essence theories with the Lagrangians of the form 
   $F(X)$, $X \equiv \phi_{,\alpha} \phi^{,\alpha}$. The instability under 
   spherically symmetric perturbations is proved for two recently obtained 
   exact solutions for  $F(X) =F_0 X^{1/3}$ and for $F(X) = F_0 X^{1/2} - 2 \Lambda$, 
   where $F_0$ and $\Lambda$ are constants. The first solution 
   describes a black hole in an asymptotically singular space-time, 
   the second one contains two horizons of infinite area connected by a wormhole.
   It is argued that spherically symmetric k-essence configurations with 
   $n < 1/2$ are generically unstable because the perturbation equation is not of hyperbolic type.}

\bigskip

] 
\email 1 {kb20@yandex.ru}
\email 2 {fabris@pq.cnpq.br}

\section{Introduction}

General Relativity (GR) theory has been very successful when tested at local scales \cite{will}. However,
GR must face, at same time, many problems. One of them is the presence of singularities in its applications to cosmology and compact objects, especially black holes. Moreover, the standard cosmological model (SCM), in spite of fitting remarkably well the observational data, requires exotic, undetected forms of matter which composes the dark sector of the universe. The SCM requires also an initial phase of accelerated expansion, the inflationary phase, in order to explain the very particular initial conditions leading to the presently observed universe. The mechanism behind this inflationary phase is still an object of debate, leading to many speculative models.

Among the modifications introduced in the GR framework in order to account for an inflationary phase in the primordial universe, a particularly interesting one is the class of k-essence modeles \cite{mukha2,mukha1} which consists in a scalar field minimally coupled to the Einstein-Hilbert term but with a nonstandard kinetic term. Such a structure can be related to more fundamental theories like string theories. One example of this connection is the Dirac-Born-Infeld action \cite{dbi}. Even if the initial motivation for the k-essence model was the inflationary cosmological phase, subsequently it has also been used also for the description of the present accelerated expansion phase of the universe \cite{mukha3}.

In a previous paper \cite{we-gc16}, we have analyzed the possible black hole type structures in the 
context of k-essence theories. The k-essence class of theories is described by the Lagrangian
\bearr 			\label{L1}
	{\cal L} = \sqrt{-g}[R - F(X,\phi)],
\ear
  with
\bear                  \label{X}
	X = \eta\phi_{;\rho}\phi^{;\rho},
\ear
$\eta = \pm 1$. In general, $F(X,\phi)$ can be any analytical function of $X$ and $\phi$.

  In Ref. \cite{we-gc16} the function $F(X,\phi)$ has been fixed as
\beq  \label{class}	
		F(X,\phi) = F_0 X^n - 2V(\phi), \quad F_0 = \const.
\eeq
  The emphasis was in obtaining complete analytical solutions. Two cases have been analyzed in detail.  
  The first one consists of a pure kinetic term, with $n = 1/3$ and $V(\phi) = 0$, leading to a 
  Schwarzschild type black hole immersed in a singular space-time, the singularity placed at spatial 
  infinity. The other solution is obtained by fixing $n = 1/2$ and $V(\phi) = \Lambda =$ constant leading to 
  a non-asymptotically flat regular space-time, with a degenerate horizon, a structure locally similar to
  the cold black holes that are present in scalar-tensor theories of gravity \cite{k1,k2}: 
  In both cases, the horizons have an infinite area and zero Hawking temperature.

  The goal of the present paper is to study the stability of those static, spherically symmetric 
  configurations found in the context of the class of k-essence theories described above. The study 
  of stability of black hole-type structures with scalar fields is a controversial subject with many
  conceptual and technical issues, see, e.g., \cite{gonz, k-z, kon, k78, turok} and references therein. 
  In performing the stability analysis for the k-essence black hole-type structures found in \cite{we-gc16},
  we closely follow the approach used in \cite{k-z}. We conclude that the solutions found
  in Ref. \cite{we-gc16} are unstable with respect to radial linear perturbations.

  The paper is organized as follows. In the next section we set up the relevant perturbed equations. In
  Section 3 we apply the perturbation analysis to the solutions found in Ref. \cite{we-gc16}, 
  and in Section 4 we present our conclusions.

\section{Perturbation equations}

  Varying the Lagrangian (\ref{L1}) with respect to the metric and
  the scalar field, we obtain the field equations
\bearr                                   		    \label{EE}
	G_\mu^\nu = -T\mN [\phi],
\yyy
	T\mN [\phi] \equiv \eta F_X \phi_\mu \phi^\nu
		- \frac{1}{2} \delta\mN F,
\yyy   					      		 \label{eq-phi}
       \eta\nabla_\alpha (F_X \phi^\alpha) - \Half\, F_\phi = 0,
\ear
  where $G\mN$ is the Einstein tensor, $F_X = \d F/\d X$, $F_\phi = \d
  F/\d\phi$, and $\phi_\mu = \d_\mu \phi$.

  We consider a \sph\ space-times with the metric 
\beq                                              \label{ds}
	ds^2 = e^{2\gamma(u,t)}dt^2 - e^{2\alpha(u,t)}du^2
			- e^{2\beta(u,t)} d\Omega^2,
\eeq
  (where $d\Omega^2 = d\theta^2 + \sin^2 \theta d\varphi^2$ is the metric on
  a unit sphere) containing only small time-dependent perturbations from 
  static ones, Accordingly, we assume 
\[
                     \alpha(u,t) = \alpha(u) + \da(u,t)
\]    
 with small $\da$, and similarly for $\beta(u,t)$ and $\gamma(u,t)$.
 The radial coordinate $u$ is arbitrary, admitting any reparametrization 
 $u \mapsto {\bar u}(u)$. The nonzero Ricci tensor components can be written 
 in the form (preserving only linear terms with respect to time derivatives)
\bear
     R^t_t \eql                                               \label{R00t}
     \e^{-2\gamma}(\ddot\alpha + 2\ddot\beta)
\nnn \cm     
           -\e^{-2\alpha}[\gamma'' +\gamma'(\gamma'-\alpha'+2\beta')],
\yy                                                            \label{R11t}
     R^u_u \eql 
     \e^{-2\gamma}\ddot\alpha                           
     - \e^{-2\alpha}[\gamma''+2\beta'' +\gamma'{}^2
\nnn \cm
     +2\beta'{}^2 - \alpha'(\gamma'+2\beta')],
\yy                                                                                                   \label{R22t}
     R^\theta_\theta \eql R^\varphi_\varphi = \e^{-2\beta} +\e^{-2\gamma}\ddot\beta
\nnn \cm    
              -\e^{-2\alpha}[\beta''+\beta'(\gamma'-\alpha'+2\beta')],
\yy
     R_{tu}\eql  2[\dot\beta' + \dot{\beta}\beta'                            \label{R01t}
                 -\dot{\alpha}\beta'-\dot{\beta}\gamma'],
\ear
  where dots and primes stand for $\d/\d t$ and $\d/\d u$, respectively.
  
  In a similar way, we assume $\phi = \phi(u,t) =  \phi(u) + \df(u,t)$ with 
  small $\df$. An accord with \eqn{X}, for static (and slightly nonstatic) scalar
  fields, to keep $X$ positive, we assume $\eta =-1$, so that, preserving only
  linear terms in $\df$, we have 
\bearr                           \label{dX} 
                  X = \e^{-2\alpha}\phi'{}^2,
\nnn                  
                  \delta X =  2\e^{-2\alpha}(\phi' \df' - \phi'{}^2 \da).
\ear        
  Then we obtain the following expressions for the nonzero stress-energy tensor
  (SET) components:
\bearr                                                       \label{SET}
	T^t_t = T^\theta_\theta = T^\varphi_\varphi = - F/2,\quad\ T^u_u = -F/2 + X F_X,
\nnn
	T_{tu} = - F_X \dot\phi \phi', 
\ear
  where, taking into account the perturbations, we should understand  $F$ as 
  $F(X) + F_X \delta X + F_\phi \df$ and $F_X$.as 
  $F_X (X) + F_{XX} \delta X + F_{X\phi} \df$. 

  In what follows, we will consider a more narrow class of k-essence Lagrangians: 
  instead of $F(X, \phi$, we take simply $F(X)$. 
  Then the scalar field equation and the nontrivial components of the Einstein
  equations can be written as follows:
\bearr                                                        \label{eq-s}
	F_X e^{\alpha + 2\beta - \gamma} {\ddot \phi} 
	         -   \Big(F_X e^{-\alpha + 2\beta + \gamma} \phi'\Big)' =0,
\yyy                                                           \label{11}
          - e^{2\alpha-\beta} + \beta'(\beta' + 2\gamma') = (\half F - X F_X)e^{2\alpha},
\yyy                                                            \label{00}
	- e^{2\alpha - 2\beta} + 2\beta'' + 3 \beta'{}^2 - 2\alpha'\beta' = \half F e^{2\alpha}, 
\yyy	
	 \e^{2\alpha -2\beta} + \e^{2\alpha -2\gamma}\ddot\beta
              - \beta'' - \beta'(\gamma'-\alpha'+2\beta')
\nnn \inch						\label{22}              
					= - \half e^{2\alpha} (F - X F_X),
\yyy
	  	\dot\beta' + \dot{\beta}\beta'                            \label{01}
                 -\dot{\alpha}\beta'-\dot{\beta}\gamma' = \half F_X \dot\phi \phi'.
\ear
  Equations \rf{11}-\rf{01} are, respectively, the equations $G^u_u= \ldots$,
  $R^t_t = \ldots$, $R^\theta_\theta = \ldots$,, and $R_{tu} = \ldots$.  
  
   Now let us suppose that the static (background, with all ``deltas'' equal to zero) 
   equations are satisfied and consider the perturbations. Since we are working 
   in the framework of GR, spherically symmetric tensor perturbations do not 
   contain an independent dynamic degree of freedom, and the only essential 
   dynamic equation  is the scalar one, \rf{eq-s}. The Einstein equations are to 
   be used in order to exclude the quantities $\da, \db, \dg$ from the scalar 
   equation. 
   
   In this problem statement, as in many similar problems \cite{},
   we have two kinds of arbitrariness, the radial coordinate choice (so far to be 
   left arbitrary), and the perturbation gauge that corresponds to fixing a 
   reference frame in perturbed space-time. As in \cite{gonz, k-z},
   it is helpful to choose the gauge $\db =0$ that substantially simplifies the 
   calculations, and, after obtaining the final form of the perturbation
   equation, it is necessary to make sure that it is gauge-invariant.   
   
   With $\db =0$, the equation for $\df$ reads 
\bearr                  \label{df1}
                - e^{2\alpha-2\gamma} \delta\ddot{\phi} + \df'' 
                   	+ \phi' \delta \sigma' + \sigma' \df'
\nnn               \inch    	
              	  + \frac {F'_X}{F_X}\df' + \phi' \delta \bigg(\frac {F'_X}{F_X}\bigg) =0,\nonumber\\[-8mm]	  
\ear
  where
\beq
                  \sigma = 2\beta+\gamma-\alpha, \cm  \delta\sigma = \dg - \da.   
\eeq    
  Since $X = e^{-2\alpha} \phi'{}^2$, from the quantities $\delta X$ and 
  $\delta X'$ we contain combinations of $\df'$, $\df''$, $\da$ and $\da'$.
  In particular, if we assume 
\beq
                  F(X) = F_0 X^n - 2 \Lambda, \qquad   n, \Lambda = \const, 
\eeq      
  equation \rf{df1} takes the form
\bearr                                    \label{df2}
	- e^{2\alpha-2\gamma} \delta\ddot{\phi} + (2n{-}1) \df'' 
			+ \df' [\sigma' - 2(n-1)\alpha'] 
\nnn \cm\cm
       	                 + \phi' [\dg' - (2n-1) \da'] = 0.
\ear  
  
  The quantities $\dg'$, $\da$ and $\da'$  can be expressed in terms of $\df$ 
  and its derivatives and the background quantities using the Einstein equations.
  More specifically, \eqn {11} gives an expression for $\dg'$ and \eqn {22} that
  for $\dg'-\da'$, containing $\da$; the latter is found from \eqn {01} as
\beq
		\da = - \frac {n}{2\beta'} X^{n-1} \phi' \df.
\eeq    
  As a result, the field equation for $\df$ takes the form
\bearr                               \label{df3}
		 -e^{2\alpha-2\gamma} \delta\ddot{\phi} + (2n{-}1) \df'' 
		 + \df' [\sigma' - 2(n-1)\alpha'] 
\nnn \cm
		- \frac{n^2}{\beta'{}^2}(e^{-2\beta}-\Lambda) e^{4\alpha} F_0 X^n \df =0.     
\ear              

  After the standard substitutions 
\bearr                    \label {u-z}
		u \mapsto z: \quad\  \frac{du}{dz} = e^{\gamma-\alpha},
\yyy                         \label {psi}		
		\df = \Psi(z)e^{i\omega t}, \qquad \Psi(z) = f(z) \psi(z), 
\nnn		
		f(z) = \exp\bigg[- \frac{\beta +(1-n)\gamma}{2n-1}\bigg],
\ear    
  we obtain a \Schr-type equation for $\psi$,
\beq                       \label{Sch}              
			(2n - 1) \frac {d^2 \psi}{dz^2} + [\omega^2 - V(z)] \psi(z) =0,
\eeq    
  where $V(z)$ is a certain effective potential for scalar perturbations, whose explicit
  form in terms of the background functions is rather bulky. 
  
  We can state that the background static solution is unstable in our linear 
  approximation if \eqn{df3} has a solution growing with time and satisfying certain
  physically relevant boundary conditions. This happens when the corresponding 
  boundary-value problem for \eqn{Sch} has solutions with $\omega^2 < 0$ since
  in this case there are physically meaningful solutions to \eqn {df3} growing as 
  $e^{|\omega| t}$.
  
  The exact boundary-value problem for \eqn{Sch} cannot be posed without invoking 
  particular background solutions. However, one general observation can be made 
  immediately. 
  
  Indeed, if $n < 1/2$, \eqn{Sch} may be rewritten as follows:
\beq
                     \frac {d^2 \psi}{dz^2} + \frac {\Omega^2 + V(z)}{1 - 2n} \psi(z) =0,
\eeq    
  where $\Omega^2 = -\omega^2$. This is a usual form of the Schr{$\ddot{\rm o}$}dinger
  equation for the ``energy level'' $\Omega^2/(1-2n)$ with the potential $W(z) =-V/(1-2n)$. 
  Depending on the particular form of $V$ and on the boundary conditions, the corresponding 
  boundary-value problem has a certain spectrum of eigenvalues, and in a majority of
  situations (in a quantum-mechanical analogy, if $-V$ does not form potential walls on both 
  ends of the $z$ range), there is a continuous spectrum  $\Omega^2 > K = \const$. Even
  if there is only a discrete spectrum, in most cases it is not bounded above. But all this
  means that $\omega^2$ is not bounded below, hence the background system is 
  catastrophically unstable and decays, in the linear approximation, infinitely rapidly --- which 
  actually means that perturbations very rapidly become large and must be considered in 
  a nonlinear mode. 
  
  Whether or not this really happens, should be verified for specific solutions and relevant 
  boundary conditions for $\df$ and $\psi$. Nevertheless, we can conclude that solutions
  with $n < 1/2$ are generically unstable.

\section{Two special solutions and their instability}
\subsection{Solution 1: $n = 1/3$, $\Lambda =0$}

 The first exact solution, obtained in \cite{we-gc16}, corresponds to the case 
 $n = 1/3$, $\Lambda =0$. It is conveniently written in terms of the so-called
 quasiglobal coordinate $u=x$ defined by the condition $\alpha(x)+\gamma(x) =0$.
 The metric has the form 
\bearr                               			\label{ds-1/3}
       ds^2 = \frac {B(x)}{k^2 x} dt^2 - \frac {k^2 x}{B(x)} dx^2 - \frac 1{k^2x} d\Omega^2,
\nnn
		B(x) = B_0 - \Half k^2 x^4,       
\ear
  where $B_0$ and $k$ are integration constants. The scalar field $\phi$ is determined 
  by the relation
\beq                  \label{phi-1/3}
		\frac {d\phi}{dx} = \phi_0 \bigg( \frac {B_0}{x^4} - \Half k^2\bigg), \qquad \phi_0 = \const. 
\eeq    
  
  If $B_0 \leq 0$, the metric function $A(x) = g_{tt} = -g^{xx} <0$, and we are dealing with
  a special case of a Kantowski-Sachs cosmological model. Therefore, in our stability
  study, we put $B_0 > 0$, in which case the metric is static at $0 < x < x_h = (2B_0)^{1/4}/k$, 
  and at $x = x_h$ there is a Killing horizon beyond which there is a cosmological 
  region. At $x=0$, where $r^2 = - g_{\theta\theta} = \infty$ (so it may be called a spatial infinity), 
  there is a singularity where both $\phi$ and the Kretschmann scalar 
  $R_{\mu\nu\rho\sigma} R^{\mu\nu\rho\sigma}$  are infinite. The Carter-Penrose diagram 
  for this space-time looks the same as for the de Sitter metric, although here a nonstatic 
  T-region corresponds to radii $r(x) < r(x_h)$, as it happens for black holes. It was therefore 
  concluded \cite{we-gc16} that this solution describes a black hole in an asymptotically 
  singular.space-time. 
  
  The perturbation equation \rf{df3} for $n=1/3$ $\alpha+\gamma =0$ and $\Lambda=0$ reads  
\bearr
		-3 e^{-4\gamma} \delta\ddot{\phi} - \df'' + 2 (3\beta'+\gamma') \df' 
\nnn \cm	\cm
		- \frac {F_0}{3 \beta'{}^2} e^{-2\beta+ 4\alpha} X^{1/3} \df =0,  
\ear      
  where, according to \rf{ds-1/3} and \rf{phi-1/3}, we should substitute    
\bearr                     \label{sub-1/3}
		e^{2\gamma} = e^{-2\alpha} = \frac{B(x)}{k^2 x}, \quad\
		\e^{2\beta} = \frac{1}{k^2 x},
\nnn
		X^{1/3} = \frac{B(x) \phi_0^{2/3}}{k^{2/3} x^3}.		
\ear  
  
  Further on, we consider a Fourier mode, $\df = \Psi(x) e^{i\omega t}$  To obtain a \Schr-like 
  equation for this mode, we first get rid of the factor $e^{-4\gamma}$ before 
  $\delta\ddot{\phi} = - \omega^2 \df$ by passing over to the ``tortoise'' coordinate $z$, such that
\beq                    \label{to_z}
		\frac {dx}{dz} = e^{\gamma-\alpha} = e^{2\gamma},
\eeq        
  which results in
\beq          \nhq
		\Psi_{zz} - 2 (3\beta_z + 2\gamma_z) \Psi_z - 3\omega^2\Psi + F_1 B(x) \Psi =0,		 
\eeq    
  where the subscript $z$ stands for $d/dz$, and $F_1 > 0$ is a constant whose particular value is
  irrelevant. Lastly, we get rid of the term with $\Psi_z$ by substituting  
\beq        \label{to_psi}
		\Psi(z) = e^{3\beta + 2\gamma} \psi(z),
\eeq    
  and with the functions \rf{sub-1/3} we obtain the equation
\bearr
		\psi_{zz} - 3\omega^2 \psi - V(z) \psi =0,
\nnn
		V(z) = - F_1 B(x) + \frac{5B_0^2}{4k^4 x^4} + \frac{31 B_0}{4k^2} - \frac{3 x^4}{16}.
\ear		  		  
  
  With this \Schr-like equation we can pose a boundary-value problem where, which is unusual,
  the role of an energy level is played by the quantity $E= -3\omega^2$. This means that if 
  the spectrum of $E$ is not restricted above, then $\omega^2$ is not restricted below, and 
  which leads to a possible growth of perturbations with arbitrarily large increments.
  
  To find out whether it is really so, let us look at the behavior of the potential at the extremes 
  of the range of $z$ and formulate the boundary conditions for $\psi$.
  
  The coordinate $x$ ranges from $x=0$ (singularity) to $x = x_h$ where $B(x) =0$ (the horizon).
  It is easy to find that $x = 0$ corresponds to a finite $z$, and we can put there $z=0$,  and then 
\beq
  	z \propto x^2\ \ {\rm as}\ \ x \to 0, \quad 
  	V(z) \approx \frac {5B_0}{4k^4 x^4} \propto .\frac 1{z^2}. 
\eeq  	
  On the other hand, near $x = x_h$,
\beq
	z \propto -\ln (x_h - x) \to \infty, \qquad  V(z) \to - \frac {8B_0}{k^2}.  
\eeq
  Thus $z \in \R_+$, and $V(z)$ is smoothly changing from an infinitely tall ``wall'' at $z=0$ 
  to a negative  constant as $z\to\infty$. 
  
  What about boundary conditions for $\psi$? Let us require that $\df$ at the boundaries 
  does not grow faster than $\phi$ itself, which, according to \rf{phi-1/3}, means that $\df$
  may grow as $x^{-3}$ near $x=0$ and should be finite as $x\to x_h$. Taking into account 
  the substitutions \rf{to_z} and \rf{to_psi}, we conclude that $\psi$ is allowed to grow
  on both boundaries not faster than  
\beq
		\psi \sim e^z \ \ {\rm as} \ \ z\to \infty, \quad \psi \sim z^{-1/4} \ \ {\rm as} \ \ z\to 0.
\eeq    

  We see that these requirements are much milder than could be imposed on a 
  quantum-mechanical wave function, therefore the spectrum of $E = -3\omega^2$ 
  is manifestly not restricted above (actually, a continuous spectrum should begin with 
  $ - 8B_0/k^2$ and extend to $+\infty$). Therefore, $\omega^2$ can take negative
  values arbitrarily large in absolute value, and thus our static configuration is 
  catastrophically unstable.
  
\subsection{Special solution 2: $n = 1/2, \Lambda > 0$}

   This solution has been obtained \cite{we-gc16} using the harmonic coordinate 
   condition \cite{kb-73}
\beq                                                   \label{harm}
  	\alpha = 2\beta + \gamma,
\eeq
  Then \eq (\ref{eq-s}) leads to
\beq
	\Big[n e^{2(1-n)\alpha} \phi'{}^{2n-1}\Big]' =0,
\eeq
   which, for $n = 1/2$, implies $\e^\alpha = a =\const$. Next, denoting $\Lambda =3/b^2$
   and choosing the length scale so that $a = b^2$, we obtain from the Einstein 
   equations \cite{we-gc16}
\bearr                                                     \label{ds-h}
	ds^2 = \frac{9}{\cosh^4 bu} dt^2 - b^4 du^2 - \frac{b^2 \cosh^2 bu}{3} d\Omega^2,
\yyy
        \sqrt X = \e^{-\alpha} |\phi'| = \frac 4{F_0}\bigg(\Lambda - \frac 2{b^2 \cosh^2 bu}\bigg)
\nnn \inch         
        			= \frac 4{F_0 b^2} (1 + 2 \tanh^2 bu),
\nnn         			
  	\phi = \pm \frac{4}{F_0 b^2}\biggl( 3 u - \frac{2}{b}\tanh bu\biggr)+ \phi_0,
\ear
  $\phi_0 = \const$. In terms of the quasiglobal coordinate $x {=}3b \tanh bu$, the metric reads
\bearr                                                       \label{ds-q}
	ds^2 =	\frac{(9b^2 - x^2)^2}{9b^4} dt^2 - \frac{9b^4}{(9b^2 - x^2)^2}dx^2
\nnn \inch
		- \frac{3b^4}{9b^2 - x^2}d\Omega^2,
\ear
  from which it is clear that there are two second-order (degenerate) horizons at $x =\pm 3b$.
 The scalar field $\phi$ in the whole range of $x$, $x \in \R$, is found as
\beq
      \phi = \pm \frac{4}{3F_0 b^4}
		\biggl(-2x + \frac{9b}{2}\ln \Big|\frac{x+3b}{x-3b}\Big|\biggr) + \phi_0
\eeq
  and is singular on the horizons $x=\pm 3b$, while $X = A\phi'^2$ is 
  infinite only as $x \to \pm \infty$ and is finite on the horizons.
  
  Considering the stability of the static region and applying the perturbation equations 
  to our case,  we notice that in \rf{ds-h} or \rf{ds-q} we have $2\beta + \gamma = 2 \ln b = \const$,
  hence $2\beta' + \gamma' = 0$, and  \eqn{df3} now takes the form 
\beq                                                            \label{df-h}
    		 \delta\ddot{\phi} = h(u) \df,     		 
\eeq  
  where 
\beq                  \label{h(u)}
		h(u) = \frac{27} {b^4 \cosh^2 bu} (1 + 2 \tanh^2 bu) > 0
\eeq   
  
  Equation \rf {df-h} is solved explicitly. Its first integral reads 
\beq
		\delta\dot{\phi}{}^2 = \df^2 h(u) + C_1 (u), \quad\ C_1(u) \ {\rm arbitrary}.
\eeq    
  Let us put $C_1(u) \equiv 0$ and suppose that $h(u) > 0$ at least in some range of $u$. 
  We then obtain 
\beq   \nhq
		\df (t,u) = e^{\pm\sqrt{h(u)} + C_2(u)},  \quad C_2(u) \ {\rm arbitrary}.
\eeq    
  Evidently, in the solution with the plus sign, $\df$ grows with time. On the other hand, 
  the arbitrariness of $C_2(u)$  makes it possible to satisfy {\it any\/} boundary conditions 
  of the form $\df \leq q(u)$ where $q(u)$ is specified on each boundary from some 
  physical requirements like finite perturbation energy, etc. 
  
  We conclude that {\it a background solution to the k-essence equations with $n=1/2$ 
  is unstable if the function $h(u)$ is positive in some range of the coordinate $u$.}  
  
  This obviously  applies to our solution since the function \rf{h(u)} is positive at all $u$, 
  that is, in the whole static region. It means that our static solution is unstable.

\section{Conclusion}

  In this work we have analyzed the stability of black hole-type configurations found previously in 
  the framework of the k-essence theory \cite{we-gc16}. These solutions were obtained in the 
  case where the k-essence function $F(X,\phi)$ is a power law given by 
  $F(X,\phi) = F_0 X^n - 2 V(\phi)$. The special cases $n = 1/3$ (with $V(\phi) = 0$) and $n = 1/2$ 
  (with $V(\phi) =\const$) admit analytical solution for a static, spherically symmetric configuration. 
  The case $n = 1/2$ leads to a consistent solution in the presence of a cosmological constant. 
  Both solutions are not asymptotically flat. For $n = 1/3$ the spatial infinity is singular, while the 
  geometry of the case $n = 1/2$ is regular, with a degenerate horizon similar to cold black holes 
  existing in scalar-tensor theories \cite{k1, k2}.

  Linear radial perturbations were considered. The analysis was performed using the gauge condition
  $\delta\beta = 0$, where $\beta$ is the logarithm of the radius of coordinate two-spheres. This 
  choice is consistent with the gauge-invariant approach for perturbations in static, spherically 
  symmetric configurations \cite{gonz,k-z,kon}. Both black hole-type solutions found in the k-essence
  framework described above turn out to be unstable: it is possible to obtain solutions for the 
  perturbed equations which grow unboundedly with time, satisfying at the same time the 
  required boundary conditions. 
  
  Moreover, it is argued that, at least generically, all \ssph\ 
  k-essence  configurations with $n < 1/2$ must be unstable: this happens, in fact, because the 
  master equation for perturbations loses its hyperbolic nature. This is, however, not proved in a 
  general form because it is necessary to take into account physically motivated boundary conditions
  for each particular solution. 
  
A point of interest is that for the intermediate value of $n$, namely, $n=1/2$ we see a rare case 
  where the perturbation equation can be directly analytically integrated without a decomposition into
  Fourier modes; this analytical solution explicitly shows the instability of the background static 
  solution.
  
  The above results may be compared, amomg others, with those reported in \cite{k-z}, where the 
  classes of scalar-vacuum static, spherically symmetric solutions known as the Fisher (ordinary 
  scalar field) and anti-Fisher (phantom scalar field) solutions, which exhibit some features similar 
  to the structures studied here, have been analyzed using a similar method. This analysis has led 
  to a conclusion on the instability of those scalar-vacuum solutions. It seems hard to make a 
  statement on the generality of the stability issue of scalar-tensor black holes, but the results 
  obtained until now may lead to some hints on this question. In particular, to our knowledge,
  there are only two examples of spherically symmetric black hole solutions with scalar fields 
  which are stable under linear spherical perturbations. One such  example is a black hole with 
  a massless conformal scalar field \cite{bbm70, bek74}, whose stability was proved in \cite{turok}.
  The other is a ``black universe'' configuration with a phantom self-interacting scalar field 
  \cite{pha1, pha2}, which proved to be stable in the case where the black hole horizon coincided
  with the minimum of the spherical radius \cite{kon}. Both examples are exceptional, while 
  in generic cases the solutions exhibit instabilities.  
  
\subsection*{Acknowledgments}

 We thank FAPES (Brazil) and CNPq (Brazil) for partial financial support.
 This study was financed in part by the Coordenação de Aperfeiçoamento de Pessoal de 
 Ni­vel Superior - Brasil (CAPES) - Finance Code 001.
 The work of KB was partly performed within the framework of the Center 
 FRPP supported by MEPhI Academic Excellence Project (contract 
 02.a03.21.0005, 27.08.2013) and partly funded by the RUDN University Program 5-100.

\end{document}